\documentstyle[epsf]{aa}
\begin{document}

%\draft
\def\be{\begin{equation}}
\def\ee{\end{equation}}
\def\bea{\begin{eqnarray}}
\def\eea{\end{eqnarray}}
\def\c{\cite}

\def\et{ {\it et al.}}
\def\la{ \langle}
\def\ra{ \rangle}
\def\ov{ \over}
\def\ep{ \epsilon}

\def\mdot{\ifmmode \dot M \else $\dot M$\fi}    % accretion rate
\def\mxd{\ifmmode \dot {M}_{x} \else $\dot {M}_{x}$\fi}
\def\med{\ifmmode \dot {M}_{Edd} \else $\dot {M}_{Edd}$\fi}
\def\bff{\ifmmode B_{f} \else $B_{f}$\fi}

\def\apj{\ifmmode ApJ \else ApJ \fi}    % lower
\def\apjl{\ifmmode  ApJ \else ApJ \fi}    %
\def\aap{\ifmmode A\&A \else A\&A\fi}    %
\def\mnras{\ifmmode MNRAS \else MNRAS \fi}    %
\def\nat{\ifmmode Nature \else Nature \fi}
\def\prl{\ifmmode Phys. Rev. Lett. \else Phys. Rev. Lett.\fi}
\def\prd{\ifmmode Phys. Rev. D. \else Phys. Rev. D.\fi}

\def\ms{\ifmmode M_{\odot} \else $M_{\odot}$\fi}    % lower
\def\na{\ifmmode \nu_{A} \else $\nu_{A}$\fi}    % Alfven frequency
\def\nk{\ifmmode \nu_{K} \else $\nu_{K}$\fi}    % Keplerian frequency
\def\ns{\ifmmode \nu_{{\rm s}} \else $\nu_{{\rm s}}$\fi}
\def\no{\ifmmode \nu_{1} \else $\nu_{1}$\fi}    % lower
\def\nt{\ifmmode \nu_{2} \else $\nu_{2}$\fi}    % upper
\def\ntk{\ifmmode \nu_{2k} \else $\nu_{2k}$\fi}    % upper
\def\dnmax{\ifmmode \Delta \nu_{max} \else $\Delta \nu_{2max}$\fi}
\def\ntmax{\ifmmode \nu_{2max} \else $\nu_{2max}$\fi}    % upper
\def\nomax{\ifmmode \nu_{1max} \else $\nu_{1max}$\fi}    % upper
\def\nn{\ifmmode \nu_{\rm NBO} \else $\nu_{\rm NBO}$\fi}    % HBO
\def\nh{\ifmmode \nu_{\rm HBO} \else $\nu_{\rm HBO}$\fi}    % HBO
\def\nqpo{\ifmmode \nu_{QPO} \else $\nu_{QPO}$\fi}    % HBO
\def\nz{\ifmmode \nu_{o} \else $\nu_{o}$\fi}    % HBO
\def\nht{\ifmmode \nu_{H2} \else $\nu_{H2}$\fi}    % HBO
\def\ns{\ifmmode \nu_{s} \else $\nu_{s}$\fi}    % stellar
\def\nb{\ifmmode \nu_{{\rm burst}} \else $\nu_{{\rm burst}}$\fi}
\def\nkm{\ifmmode \nu_{km} \else $\nu_{km}$\fi}    % stellar
\def\ka{\ifmmode \kappa \else \kappa\fi}    % stellar
\def\dn{\ifmmode \Delta\nu \else \Delta\nu\fi}    % stellar

\def\vk{\ifmmode v_{k} \else $v_{k}$\fi}    % Keplerian velocity
\def\va{\ifmmode v_{A} \else $v_{A}$\fi}    % Alfven velocity
\def\vf{\ifmmode v_{ff} \else $v_{ff}$\fi}    % free fall velocity

\def\rs{\ifmmode R_{s} \else $R_{s}$\fi}    % stellar
\def\ra{\ifmmode R_{A} \else $R_{A}$\fi}    % Alfven radius
\def\rso{\ifmmode R_{S1} \else $R_{S1}$\fi}    % sonic point radius
\def\rst{\ifmmode R_{S2} \else $R_{S2}$\fi}    % sonic point radius
\def\rmm{\ifmmode R_{M} \else $R_{M}$\fi}    % stellar
\def\rco{\ifmmode R_{co} \else $R_{co}$\fi}    % stellar
\def\ris{\ifmmode {\rm R}_{{\rm ISCO}} \else $ {\rm R}_{{\rm ISCO}} $\fi}
\def\rsix{\ifmmode R_{6} \else $R_{6}$\fi}    % stellar

\title{ The MHD Alfven wave oscillation model of kHz Quasi Periodic Oscillations of Accreting X-ray Binaries}
\author{Chengmin Zhang}

\institute{1. National Astronomical Observatories,
 Chinese Academy of Sciences, 
Beijing 100012, China, zhangcm@bao.ac.cn\\
2. Research Center for     Theoretical Astrophysics,
School of Physics,       The University of Sydney,
       NSW 2006, Australia}
%\\ zhangcm@physics.usyd.edu.au}
%       Tel: 61-2-9351 2546
%       FAX: 61-2-9351 7726

\date{Received date ; accepted date}

\offprints{Chengmin Zhang}%,  zhangcm@bao.ac.cn}

%\begin{abstract}
\abstract{ We ascribe  the interpretation of the twin kilohertz
Quasi Periodic Oscillations (kHz QPOs) of X-ray spectra of  Low
Mass X-Ray Binaries (LMXBs) to MHD Alfven wave oscillations in
the different mass density regions of the accreted matter at the
 preferred   radius, and the upper kHz QPO frequency
coincides with the Keplerian frequency. The proposed model
concludes that the kHz QPO frequencies depend inversely on the
preferred  radius,
 and that theoretical relation between the upper frequency ($\nt$) and
 the lower frequency ($\no$) is $\no \sim \nt^{2}$, which is  similar
to the measured empirical relation.   The separation between the
twin frequencies decreases (increases) with increasing  kHz QPO
frequency if the lower kHz QPO frequency is more (less) than $\sim$
400 Hz. %\end{abstract}
}
\maketitle

\section{Introduction}

The launch of the X-ray timing satellite, Rossi X-ray Timing Explorer (RXTE),
led to the discovery of Quasi
Periodic Oscillations (QPOs) of LMXBs in their  X-ray brightness, with frequencies
$\sim 10^{-1} - 10^{3}$ Hz (see van der Klis 2000 for a recent review).
 Thereafter, much attention has been paid to the kHz QPO mechanism of LMXB; 
 however the proposed
models are still far from explaining  all detected data. The Z
sources (Atoll sources), which are high (less) luminous  neutron star (NS) LMXBs
(Hasinger \& van der Klis 1989), typically show four distinct
types of QPOs (van der Klis 2000). At present, these are the
normal branch oscillation (NBO) $\nn \simeq 5-20$~Hz, the horizontal
branch oscillation (HBO) $\nh \simeq 10-70$~Hz, and the kHz QPOs
$\nt(\no) \simeq 300-1300$~Hz that typically occur in pairs in
more than 20  sources, with upper frequency $\nt$ and lower frequency
$\no$. In
  11  sources, nearly coherent burst oscillations $\nb \simeq
270-620$~Hz
 have also
been detected during thermonuclear Type~I X-ray bursts; these are
considered to be  the NS spin frequencies $\ns$ or their first overtone (see,
e.g., Strohmayer \& 
 Bildsten 2003). Moreover, the existence of a third kHz QPO
has also been reported
 in three low-luminosity sources (Jonker et al. 2000).
  All of these QPOs but the burst oscillations have
centroid frequencies that increase with the inferred mass
accretion rate \mdot.
Furthermore, the frequencies  $\nt$ and $\no$,
 as well as the frequencies  $\nt$ and $\nh$,
 follow  very
similar relations in five Z sources, and the QPO frequencies  of
LMXBs and black hole candidates (BHC) have  a tight and
systematical correlation over three orders of magnitude in
frequency  (Psaltis et al. 1998, 1999;  Belloni et al. 2002).

Various  theoretical models have been proposed to account for
the  QPO phenomenon  in X-ray binaries (for a review see, e.g., Psaltis
2000). In the early detection of RXTE, the upper kHz QPO ($\nt$) was
simply considered to originate
 from the Keplerian orbital frequency at the preferred  radius, 
 and the lower kHz QPO
 ($\no$) is  attributed to
the beat of this  frequency with the stellar spin frequency $\ns$
 (Strohmayer et al. 1996; Miller et al. 1998). However, this beat 
model is inadequate, for the detected  frequency separation 
($\dn \equiv  \nt - \no$)   decreased  systematically
with instantaneous \mdot{} (see, e.g., van der Klis 2000). 
 Later on, general
relativistic effects were  invoked to account for kHz QPOs (Stella
\& Vietri 1999; Stella et al. 1999;  Psaltis \&
Norman 2000), which can satisfactorily explain the variation in kHz QPO
separation $\dn$.
  Moreover, the theory of epicyclic parametric resonance
 in relativistic accretion disks 
was proposed (Abramowicz et al. 2003), where the twin kHz  QPOs occur
 at the frequency of meridional
oscillation and the radial epicyclic frequency in the same orbit, 
which can explain 
the frequency ratio 3:2 detected
in black hole candidates.
  Although many other feasible ideas have  also proposed, such as
the disk seismic model (Wagoner 1999), a two-oscillator model
(Osherovich \& Titarchuk) and the
 photon bubble model (Klein et al. 1996),
 no model has yet explained satisfactorily 
 all observed  QPO  phenomena of LMXBs until now. 

In this paper, the  MHD Alfven wave oscillation model is proposed,
 and its predictions and comparisons with the 
well detected
sample sources are shown  in the  figures.
%The speed of light c and Newton's gravitational constant G are used as usual.

\section{The Model}
  The model's idea  is traced to the
  analogies with  magnetic  waves in  coronal oscillations in solar physics,
where the  MHD turbulence 
driven by Alfven wave  oscillations
occurs in the magnetic loops of  the Sun's coronal atmosphere (Roberts 2000).
 While the dynamical  details of the mechanisms
responsible for the kHz QPOs in LMXBs are still uncertain, 
it is convenient  to imagine that  the MHD Alfven wave
oscillations occur  at a certain preferred radius, where a  MHD tube
loop may be formed to conduct  
the accreted matter to the polar cap  of star.  Nevertheless, 
 it is  assumed  that this preferred  radius is a critical or a transitional
 radius  where the spherical accretion matter with low mass density
 is transferred into polar channel  accretion with  
 high mass density  that  follows  the loop and accretes onto 
 NS polar cap. Possibly, 
 this  critical  transition may  give rise to MHD
 turbulence so that  much more energy is liberated than   at other positions.

As a phenomenological prescription, we associate the twin 
kHz QPO frequencies  
with the Alfven wave oscillation frequencies (AWOFs)
at a preferred radius described in the Appendix, where
  the AWOF
 with the spherical accretion mass density coincides with
 the Keplerian orbital frequency $\nk$
and the AWOF with the polar accretion mass density 
corresponds to a 
lower frequency.  As a preliminary investigation,
 we are not concerned with the 
 actual mechanisms of  producing  the  kHz QPOs in the X-ray fluxes of
 LMXBs (see, e.g., Miller et al. 1998;
 Psaltis 2000). Rather, our main purpose is to
 stress the consistence between the model and the measured kHz QPO data, 
 and leave the arguments concerning the possibility of the mechanism to a later paper.

At the preferred  radius r, the Alfven
velocity is, $ v_{A}(r) = \frac{B(r)}{\sqrt{4\pi
\rho}}$,  where $B(r)=B_{s}(R/r)^{3}$ and $B_{s}$ are the
dipole magnetic field strengths at radius r and at the surface
of star with radius R,   respectively. Therefore,
 the AWOF  $\nu_{A}$ is 
 \be \nu_{A}(r) =  \frac{v_{A}(r)}{2\pi r } =
%\frac{B_{s}(R/r)^{3}}{2\pi r \sqrt{4\pi \rho}} =
\frac{B_{s}(R/r)^{3}}{2\pi r}  \sqrt{{Sv_{ff} \over  4\pi \mdot}}
\propto \sqrt{S} \, , \label{nua} \ee where the  mass density of
the accreted matter,  $\rho=\mdot/[S\vf]$ (Shapiro \& Teukolsky
1983) is
 applied with   the free fall
velocity $\vf = \sqrt{2GM/r} = c\sqrt{\rs/r}$, where
$\rs=2GM/c^{2}$ is the Schwarzschild radius and can be expressed as
$\rs \simeq 3m ({\rm km}) = 0.3m ({\rm 10 km}) $
 with $m={M \over \ms}$, the NS mass  in  units  of solar mass, and 
the area  S representing   the  spherical  area  
$S_{r}$  or   the polar cap area $S_{p}$, respectively,

\be
S_{r}=4\pi r^{2} \, ,
\ee

\be S_{p}=4\pi R^{2} (1 - cos\theta_{c})\, , \,\,
sin^{2}\theta_{c}={R/r}\equiv X \, , \ee where $\theta_{c}$ is the
open angle of the last field line to close at radius r. As an
approximation, the polar cap area is usually  written as
$S_{p}=\frac{2\pi R^{3}}{r}$ if $R \ll r$
 (Shapiro \& Teukolsky 1983).
  For simplicity, it is
convenient to write the two areas by means of  the scaled radius
parameter $X \equiv R/r$, so we have, \be S_{r}=4\pi R^{2} X^{-2}\, ,\,\,
S_{p}=4\pi R^{2} (1 - \sqrt{1 - X})\, . \ee
 It is assumed that $\nt$ and  $\no$ 
 are from the MHD Alfven wave  oscillations with the different accreted
material mass densities, corresponding to the different areas
 $S_{r}$ and  $S_{p}$ respectively, and $\nt$ coincides  with the
Keplerian frequency at the preferred radius (see Appendix),  
therefore, \be \nt =
\sqrt{{GM\over 4\pi^{2}r^{3}}}=\nu_{A}(S_{r}) = 1850 ({\rm
Hz}){}A{}X^{3/2}\,, \label{nt} \ee with  the parameter A defined
as $A = (m/R_{6}^{3})^{1/2}$ and  $R_{6}=R/10^{6}cm$. 
 The physical meaning of A is clear in that  $A^{2}$ represents  the
 NS averaged mass density. So, by
means of  Eq.(\ref{nua}) with the correlation $\nu_{A} \propto
\sqrt{S}$,  we obtain  the lower kHz QPO frequency, \be \no =
\nu_{A}(S_{p}) = \nt \sqrt{S_{p}/S_{r}} =  \nt X\sqrt{1 -
\sqrt{1-X}}\,. \label{no} \ee The twin kHz QPO frequencies only
depend on  two variables, the parameter A and  the scaled
radius X, so these two variables are  implied if the
twin kHz QPO  frequencies are  detected simultaneously. Therefore,
the ratio of the twin frequencies is obtained easily, by setting
$y(X)={\nt \over \no}$, \be {\no \over \nt} =  y^{-1}(X) =
X\sqrt{1 - \sqrt{1-X}}\,, \label{ratio} \ee which is independent
of  the parameter A  and is only related to the 
parameter X. Furthermore, the twin frequency separation is
given as follows, \be \Delta \nu  \equiv   \nt - \no = \nt (1 -
X\sqrt{1 - \sqrt{1-X}})\,. \label{separation} \ee

  The comparisons of the model's conclusions to the four detected 
sample  sources, Sco X-1 (van der 
Klis et al. 1997), 4U1608$-$52
(M\'endez et al. 1998a,b), 4U1735-44 (Ford et al. 1998)
 and 4U1728$-$34 (M\'endez \& van der Klis 1999), are shown in the figures.
 The relations $\nt$ versus  $\no$ and $\dn$ versus  $\no$  are  plotted in
Fig.1 and Fig.2, respectively,  and they show that the agreement  between the model
and the observed QPO data is  quite good for the   selected
ranges of  NS parameters A=0.6, 0.7 and 0.8. In Fig.2, we find
that $\dn$ increases with increasing  $\no$ if  $\no < 408$
(A/0.7) Hz  and $\dn$ decreases with increasing  $\no$ if $\no
> 408$ (A/0.7) Hz. 
 The theoretical relation between the twin frequencies is derived from
Eq.(\ref{nt}) and Eq.(\ref{no}),
\be
 \no = 629 (Hz) A^{-2/3} \nu_{2k}^{5/3}
 \sqrt{1 -\sqrt{1-({\nu_{2k} \over 1.85A})^{2/3} }}\, ,
 \label{n1n2-1}
\ee where $ \nu_{2k} = \nt/({\rm 1 kHz})$. If $\nt \ll 1850{}A
({\rm Hz}) = 1295({A\over 0.7}) ({\rm Hz})$, then  we obtain  the
approximated theoretical relation between the twin kHz QPO frequencies, 
\be
 \no = (382/A){} ({\rm Hz}){} \nu_{2k}^{2}= 546 ({\rm Hz})
 ({A \over 0.7})^{-1}  (\nu_{2k})^{2}\,.
\ee  A similar $\no$-$\nt$ empirical correlation has also been
found for the measured  kHz QPO  sources (see, e.g., Stella et al.
1999; Psaltis et al. 1998; Psaltis 2000; Psaltis and Norman 2000).
 In Fig.3, we plot the $\nt/\no$ versus $\no$ diagram, and it is found that 
the ratio $\nt/\no$
decreases with increasing  $\no$. The averaged frequency
ratio for the four sample sources is about $<y>=1.4=7:5$,
corresponding  to the averaged X, $<X>=0.88$.
 In Fig.4, we plot  $\nt/\no$ versus X, and find that
 the theoretical  curve is independent of the parameter A, which
 reflects   the pure
 geometrical scaling relation  between the twin frequency
ratio and the parameter X=R/r. The X distributions for the four
examples are very similar (from X=0.8 to X=0.93),
which implies that the dynamic mechanism
 that accounts for kHz QPO is intimately related to the scaled radius
X and has no direct relation with the other physical quantities.

%\section{Discussions and Conclusions}
In conclusion,  the consistence between the model and the detections is robust, and 
the main results are summaried as follows: 
 (1) the twin QPOs  are inversely (proportionally) related to the  radii 
(the accretion rate); (2) the theoretical relation between the twin frequencies is 
$\no \sim \nt^{2}$; (3) the separation between the
twin frequencies decreases (increases) with increasing the kHz QPO
frequency if the lower kHz frequency is more (less) than $\sim$
400 Hz; (4) the ratio between the  twin frequencies is  only related to the scaled 
radius parameter X, and the homogenous kHz QPO frequency distributions of 
four detected sources indicate that these frequencies are from  regions 
close to  the  surface of the NS, approximately, from 1.08R to 1.25R.  
 With regard to  conclusion (3), for the theoretical $\dn$-$\no$ relation with 
$\no < 400$ Hz future detection is needed to confirm  this prediction.

\vskip .3cm
\begin{acknowledgements}
Thanks are due to T. Belloni, M. M\'endez, and D. Psaltis for providing
the data files, and discussion  with T.P. Li is highly appreciated.
\end{acknowledgements}

%\newpage

{\bf APPENDIX: On the preferred radius}\\

The Alfven radius $\ra$ is determined by the condition that 
the magnetic ram pressure
should match the plasma momentum energy density (Shapiro \& Teukolsky 1983),
 where  $ \ra = 1.9\times 10^{6} (cm)
B_{s8}^{4/7}\mdot_{17}^{-2/7}m^{-1/7}R_{6}^{12/7}$,
and $B_{s8}$ the surface field in unit of $10^{8}$ ~ Gauss,
$\mdot_{17}$ the accretion rate in unit of $10^{17}$ ~ g/s,

\be
{B^{2}(\ra)\over 8\pi} = {1\over 2}\rho(\ra)v^{2}(\ra)\,,
\ee
or equivalently, $\va(\ra)={B(\ra)\over \sqrt{4\pi\rho(\ra)}} = v(\ra)$, 
where $\va$ and $v$ represent the Alfven velocity and the free fall velocity,  
$v(r)=v_{ff}(r)=\sqrt{2GM/r} =c\sqrt{\rs/r}=\sqrt{2}\vk(r)$
 and $\rho(r)=\rho_{ff}(r)=\mdot/[S\vf(r)]$ with the 
area factor $S = 4\pi r^{2}$.    In other 
words, the Alfven velocity equals  the free fall
velocity  at the  Alfven radius.  
  If the area factor S decreases 
(the polar cap area for instance), the mass density $\rho$ will increase and 
 the Alfven velocity will decrease.
 If we set $v(r)=\vk(r)$ and $\rho(r)$ = $ \rho_{ff}(r)$,
 then  we will obtain a preferred  radius where the Alfven velocity
matches the Keplerian flow velocity,

\be
\va(\rso)={B(\rso)\over \sqrt{4\pi\rho(\rso)}} = \vk(\rso)\,.
\label{vso}
\ee
So, the radius $\rso$ can be obtained to be  $\rso=2^{2/7}\ra=1.22\ra 
\sim (B_{s}/\mdot^{1/2})^{4/7}$, 
where the Alfven wave oscillation frequency matches the Keplerian 
frequency,
$\na(\rso)=\nk(\rso)$; we prefer to call  it 
the quasi sonic-point radius as a distinction
of the  sonic-point radius  by  Miller et al. (1998) and Lai (1998).

%\newpage
%\clearpage
\begin{figure}
\epsfxsize=4.5cm
\includegraphics{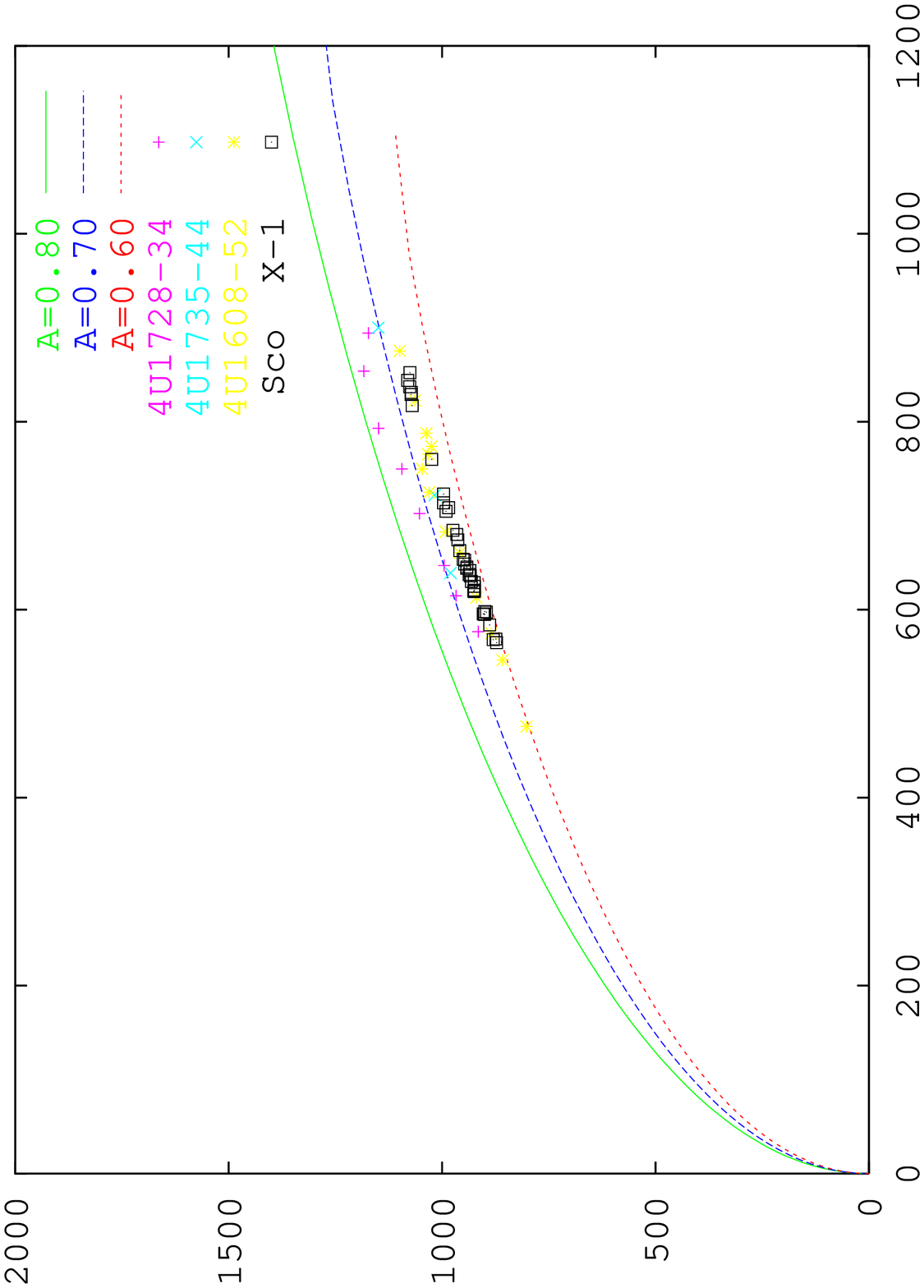}
\vskip 6.0cm
\caption[fig1] {$\nt$ versus  $\no$ plot.
The horizontal axis is the lower kHz QPO frequency $\no$
and  the vertical axis is  the upper kHz QPO frequency $\nt$.
 The  kHz QPO data of four detected sample  sources are plotted. The model
presents a good  consistence  with the measured data for  the NS averaged mass density
parameters  A=0.6, 0.7 and 0.8, which are  shown in the three theoretical curves
 from bottom to  top.}\label{fig1}
\end{figure}

\begin{figure}
\epsfxsize=4.5cm
{\includegraphics{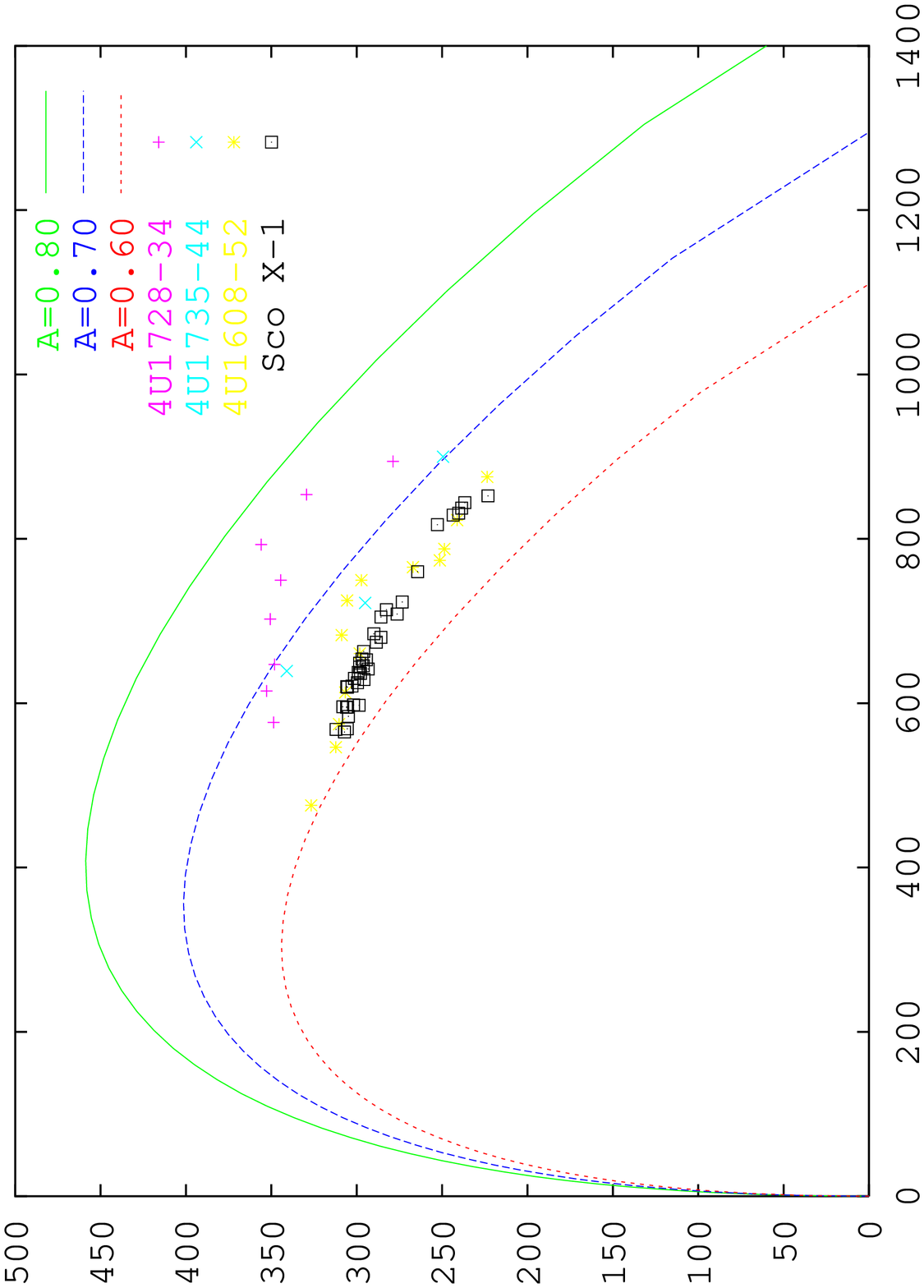}}
\vskip 6.0cm
\caption[fig2] {$\dn$ versus  $\no$ plot.
The horizontal axis is the lower kHz QPO frequency $\no$
and  the vertical axis is  the twin kHz QPO frequency separation $\dn$.
 The  kHz QPO data of four detected sample  sources are plotted.
The model
presents a good  consistence  with the measured data for  the NS averaged mass density
parameters  A=0.6, 0.7 and 0.8, which are  shown in the three theoretical curves
 from bottom to  top.}\label{fig2}
\end{figure}

\begin{figure}
\epsfxsize=4.5cm
{\includegraphics{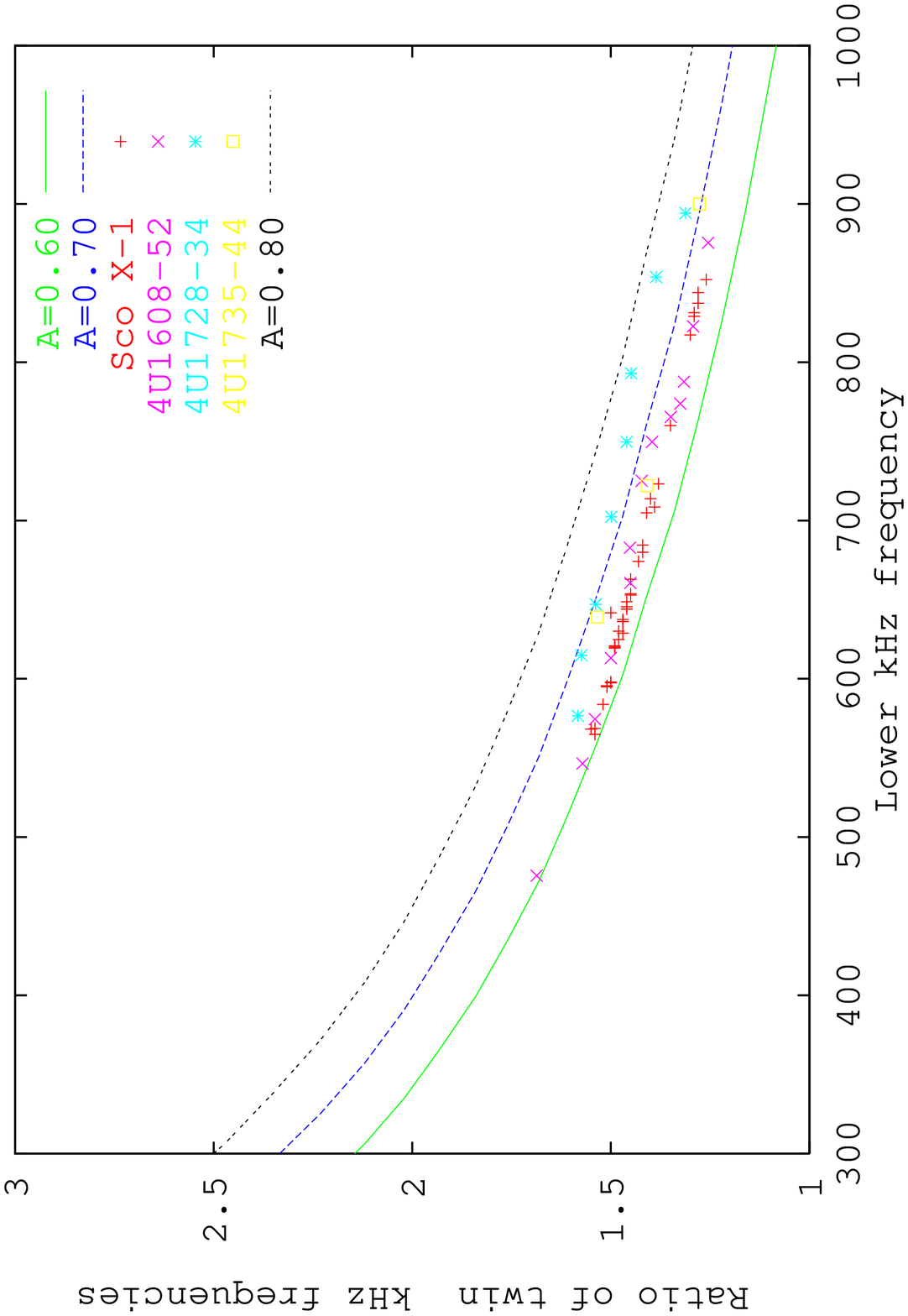}}
\vskip 6.0cm
\caption[fig3] {The ratio between the  twin kHz  QPO frequencies  versus the
 lower kHz QPO frequency.  The three
 theoretical curves represent  the NS mass density parameter conditions
with A=0.6, 0.7 and 0.8 from bottom to  top.
The data of four sample  sources are plotted.} \label{fig3}
\end{figure}

\begin{figure}
\epsfxsize=4.5cm
{\includegraphics{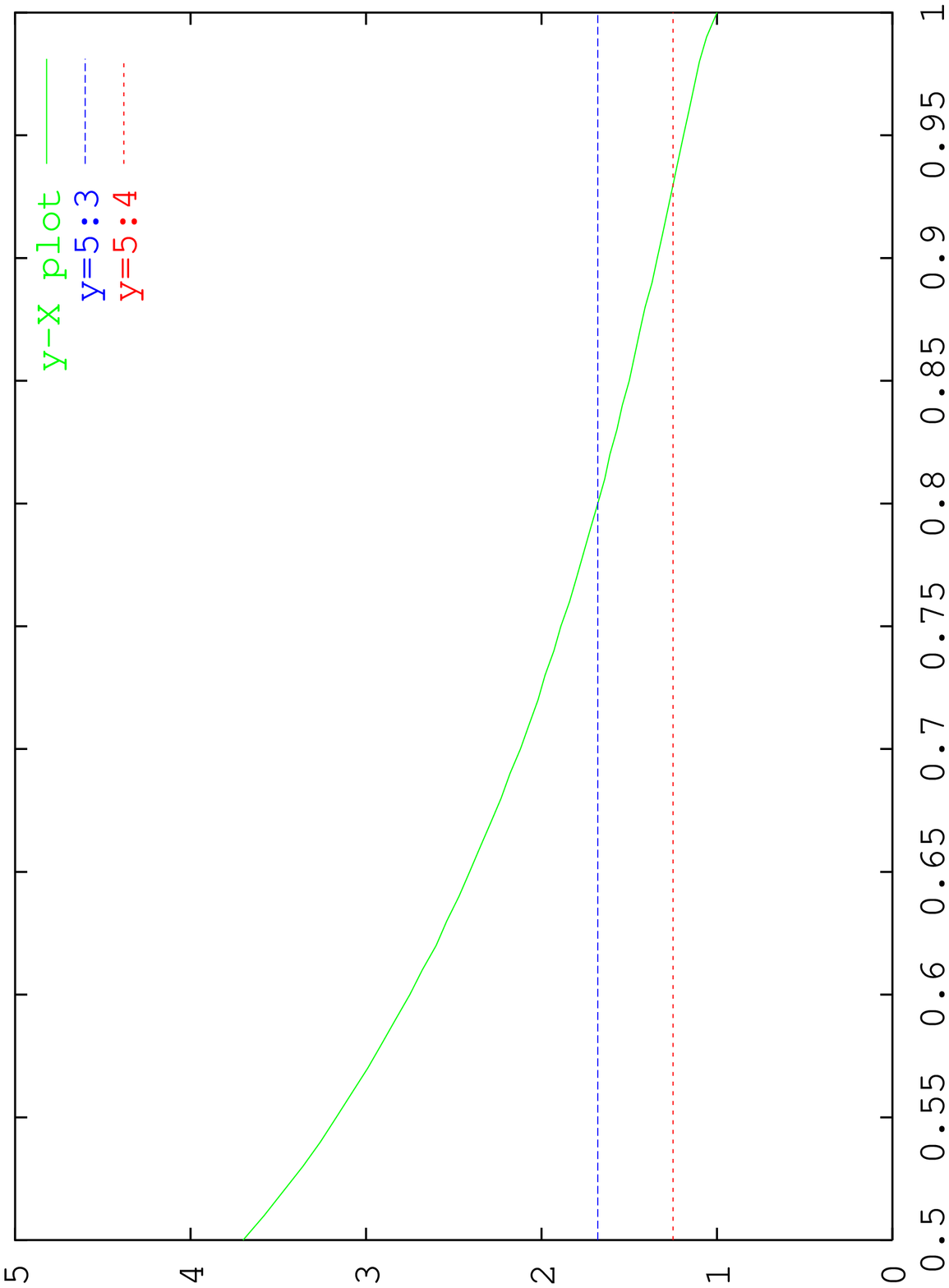}}
\vskip 6.0cm
\caption[fig4] {The ratio between the twin kHz  QPO frequencies 
    versus the parameter $X\equiv R/r$.
The horizontal axis is X
and  the vertical axis is  the ratio y=$\nt/\no$.
 It is found that the  theoretical curve is independent of the NS
mass density parameter A. 
The two horizontal  lines represent y=5:4 and y=5:3,
from bottom to top, respectively,
 which cover  the ratios of the  
detected kHz QPO data of the four samples.
} \label{fig4}
\end{figure}

\end{document}